\documentclass[aps,prl,twocolumn,superscriptaddress,longbibliography]{revtex4-2}
\usepackage{graphicx}
\usepackage{amssymb,amsmath}
\usepackage{bm}
\usepackage{dcolumn}
\usepackage{float}
\usepackage[OT1]{fontenc} 
\usepackage{url}
\usepackage{mathrsfs}
\usepackage{slashed,comment}
\usepackage{color}
\usepackage{verbatim}
\usepackage{enumitem}
\usepackage{soul,physics}
\usepackage[driverfallback=dvipdfm]{hyperref}
\hypersetup{pdfpagemode=FullScreen,colorlinks=true,breaklinks,urlcolor=blue,linkcolor=blue,citecolor=blue}

\usepackage{subfigure}
\usepackage{amssymb}
\usepackage{bm}
\usepackage{graphicx}
\usepackage{amsmath,amssymb}

\begin{document}
 
  \title{Many-Body Physics from Spin-Phonon Coupling in Rydberg Atom Arrays}

  \author{Shuo Zhang}
  \affiliation{Department of Physics, Fudan University, Shanghai, 200438, China}

  \author{Langxuan Chen}
  \affiliation{Department of Physics, Fudan University, Shanghai, 200438, China}

  \author{Pengfei Zhang}
  \thanks{PengfeiZhang.physics@gmail.com}
  \affiliation{Department of Physics, Fudan University, Shanghai, 200438, China}
  \affiliation{State Key Laboratory of Surface Physics, Fudan University, Shanghai, 200438, China}
  \affiliation{Hefei National Laboratory, Hefei 230088, China}

  \date{\today}

  \begin{abstract}
  The rapid advancement of quantum science and technology has established Rydberg atom arrays as a premier platform for exploring quantum many-body physics with exceptional precision and controllability. Traditionally, each atom is modeled as a spin degree of freedom with its spatial motion effectively frozen. This simplification has facilitated the discovery of a rich variety of novel equilibrium and non-equilibrium phases, including $\mathbb{Z}_{\text{N}}$ symmetry-breaking orders and quantum scars. In this work, we investigate the consequences of incorporating atomic vibrations in optical tweezers, which give rise to spin-phonon coupling. For systems in thermal equilibrium, we find that this coupling leads to a new symmetry-breaking phase in the weak driving limit, as a result of induced three-spin interactions. Furthermore, we show that the violation of quantum thermalization in $\mathbb{Z}_2$-ordered states is suppressed when spin-phonon coupling is introduced. Our results are readily testable in state-of-the-art Rydberg atom array experiments.
  \end{abstract}
    
  \maketitle

  \emph{ \color{blue}Introduction.--} 
  Understanding novel quantum phases of many-body systems is a central topic in quantum many-body physics. In thermal equilibrium, the traditional framework proposed by Landau classifies phases of matter based on their symmetry-breaking patterns~\cite{landau}, leading to conventional phases such as superconductivity and magnetism. Subsequent developments have revealed that the interplay between symmetry and topology in many-body wave functions plays a crucial role in characterizing more exotic phases~\cite{RevModPhys.89.041004,RevModPhys.88.035005}. Beyond equilibrium, quantum systems can also be classified into distinct phases based on their dynamical properties. For example, generic Hamiltonians are expected to exhibit quantum thermalization~\cite{PhysRevA.43.2046,PhysRevE.50.888,Rigol:2007juv,DAlessio:2015qtq}, while strong disorder may prevent thermalization, leading to a transition into a many-body localized phase~\cite{RevModPhys.91.021001,2018CRPhy..19..498A,2015ARCMP...6..383A,Nandkishore:2014kca,2013PhRvL.111l7201S,Huse_2014,2013PhRvL.110f7204V,2015Sci...349..842S,2016Sci...352.1547C,2016NatPh..12..907S}. Moreover, recent studies have identified quantum many-body scars, where nonthermal states are embedded within an otherwise thermalizing spectrum~\cite{Turner:2018yco,Turner_2018,PhysRevLett.122.040603,PhysRevLett.122.220603,PhysRevX.10.021041,PhysRevX.11.021021,Serbyn:2020wys,PhysRevB.98.235155,PhysRevB.101.165139,2019PhRvL.122q3401L,2020PhRvB.101i4308M,Mark:2020yjs,Lin:2019qlo,2020PhRvB.102g5132M,PhysRevResearch.4.L032037,Cheng:2024nep}.

  The development of quantum science and technology has brought new opportunities for observing novel quantum phases in experimental platforms with high controllability. In particular, Rydberg atom arrays have emerged as a key platform~\cite{RevModPhys.82.2313,2010NatPh...6..382W,Labuhn_2016,Bernien:2017ubn,2020NatPh..16..132B,Ebadi:2020ldi,Scholl:2020hzx,Wu:2020axb,Semeghini:2021wls,fang2024probingcriticalphenomenaopen,Cheng:2024nep,Bluvstein_2021,Morgado:2020jfo,Bluvstein:2021jsq,Ma:2023ltx,Scholl:2023cjt,Singh:2022qfv,Bluvstein:2023zmt,zhang2025observationnearcriticalkibblezurekscaling}, where both equilibrium phases such as $\mathbb{Z}_{\text{N}}$ symmetry-breaking orders~\cite{Bernien:2017ubn,2020NatPh..16..132B,Ebadi:2020ldi,Scholl:2020hzx} and non-equilibrium phases including quantum scars~\cite{Turner_2018,Bernien:2017ubn,Bluvstein_2021} have been observed experimentally. In traditional theoretical descriptions, each atom trapped in an optical tweezer is modeled as a quantum spin, with its spatial motion effectively frozen. However, recent developments have uncovered novel phenomena arising from the coupling between spin and spatial motion, effectively forming a spin-phonon interaction~\cite{P_odzie__2018,PhysRevLett.125.033602,Magoni:2021jgi,PhysRevLett.131.093002}. This coupling gives rise to polaronic quasiparticle excitations and the emergence of the Jahn-Teller effect~\cite{1937RSPSA.161..220J,10.1098/rspa.1938.0008} in few-body systems. Despite these advances, the influence of spin-phonon coupling on many-body equilibrium and non-equilibrium phases remains largely unexplored.

  \begin{figure}[t]
    \centering
    \includegraphics[width=0.9\linewidth]{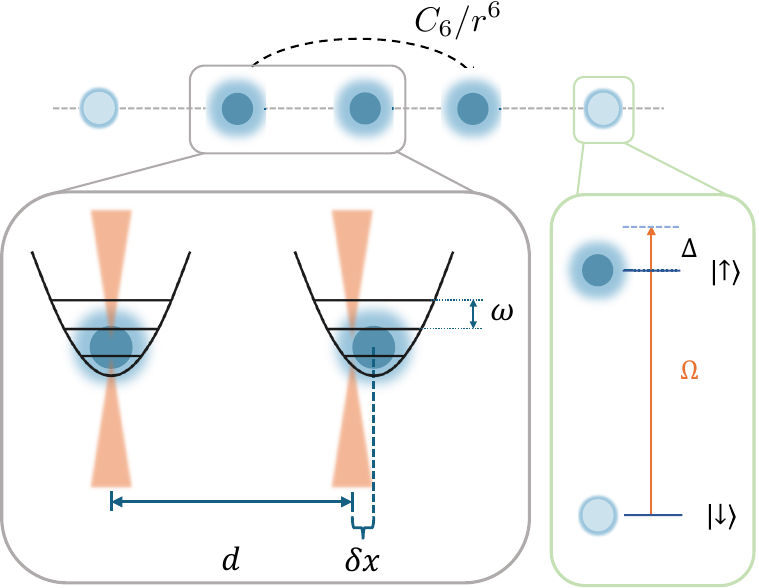}
    \caption{We present a schematic of the model considered in this manuscript. The system consists of a one-dimensional array of optical tweezers, each trapping a single atom. The internal state of each atom is described by a quantum spin, with $\ket{\uparrow}$ and $\ket{\downarrow}$ denoting the Rydberg and ground states, respectively. The spatial motion of each atom is modeled as a quantum harmonic oscillator. The two internal states are coherently coupled with an effective Rabi frequency $\Omega$ and detuning $\Delta$. Additionally, atoms in the Rydberg state interact via van der Waals interactions.  }
    \label{fig:schemticas}
  \end{figure}

  In this work, we take an initial step toward understanding how spin-phonon coupling influences both symmetry-breaking patterns and quantum scars in one-dimensional Rydberg atom arrays. Focusing on the regime of weak Rabi coupling, we derive effective three-spin interactions mediated by phonons. We begin by investigating the system’s equilibrium properties, mapping out the phase diagram as a function of detuning and Rydberg blockade radius under varying strengths of spin-phonon interaction. Notably, we identify a new $\mathbb{Z}_3$ symmetry-breaking phase that does not arise in the absence of spin-phonon coupling. We then turn to the non-equilibrium dynamics, examining the time evolution from a $\mathbb{Z}_2$ scar state, which typically exhibits persistent oscillations. Since spin-phonon coupling introduces additional decay channels, it suppresses the violation of quantum thermalization. Our results offer experimentally relevant insights into the impact of motional degrees of freedom in Rydberg-based quantum simulators.

  \emph{ \color{blue}Model.--} Our system consists of a one-dimensional lattice of optical tweezers, each trapping a single atom, as illustrated in Fig.~\ref{fig:schemticas}. Each atom possesses an internal degree of freedom associated with its electronic state, and a motional degree of freedom. The internal degree of freedom is modeled as a two-level system, where the electronic ground state $\ket{\downarrow}$ is coherently coupled to the Rydberg state $\ket{\uparrow}$, with effective Rabi frequency $\Omega$ and detuning $\Delta$. The motional degree of freedom is treated as a quantum harmonic oscillator. We assume that the trapping frequency is independent of the internal state, which can be achieved by operating the trapping laser at the magic wavelength~\cite{PhysRevA.84.043408,PhysRevA.98.033411,PhysRevLett.124.023201,Madjarov:2020nhe,PhysRevLett.128.033201}. Additionally, when two atoms at sites $i$ and $j$ are both in the Rydberg state, they experience a van der Waals interaction given by $V(|i-j|) = C_6 / |\hat{\mathbf{r}}_i - \hat{\mathbf{r}}_j|^6$, where $\hat{\mathbf{r}}_j=(\hat{x}_j,\hat{y}_j,\hat{z}_j)$ denotes the position operator of the $j$-th atom. Putting all these ingredients together, the Hamiltonian of the system reads:
  \begin{equation}
  \begin{aligned}
    \hat{H} = &\sum_{j=1}^{N} \Bigg(\frac{\Omega}{2} \hat{\sigma}^x_j -\Delta \hat{n}_j+\frac{\hat{\mathbf{p}}^2_j}{2m} + \frac{1}{2}m\omega^2\delta \hat{\mathbf{r}}^2_j\Bigg) \\&+\sum_{i<j}\frac{C_6}{|\hat{\mathbf{r}}_i - \hat{\mathbf{r}}_j|^6}\hat{n}_i \hat{n}_j ,
  \label{new hamiltonian}
  \end{aligned}
  \end{equation}
  where $N$ denotes the system size. The operator $\hat{n}_j = (\hat{\sigma}^z_j + 1)/2$ represents the number operator for the Rydberg state, and $\hat{\mathbf{p}}_j=(\hat{p}_{x,j},\hat{p}_{y,j},\hat{p}_{z,j})$ is the momentum operator of the $j$-th atom. We define the displacement operator as $\delta \hat{\mathbf{r}}_j = \hat{\mathbf{r}}_j - \mathbf{r}^0_j=(\delta \hat x_j,\delta \hat y_j,\delta \hat z_j)$, where $\mathbf{r}^0_j = (j d, 0, 0)$ denotes the center position of the $j$-th optical tweezer, with $d$ being the lattice constant. 

 We work in the regime where the displacement of each atom is much smaller than the lattice constant. As a result, we expand the interaction to leading order in $\delta\hat{\mathbf{r}}_j/d$. For $i>j$, this leads to~\cite{PhysRevLett.125.033602,Magoni:2021jgi,PhysRevLett.131.093002}
 \begin{equation}
 \begin{aligned}
    \frac{C_6}{|\hat{\mathbf{r}}_i - \hat{\mathbf{r}}_j|^6} \hat{n}_i \hat{n}_j\approx \left(\frac{C_6}{[d(i-j)]^6}-\frac{6C_6(\delta \hat{x}_i-\delta \hat{x}_j)}{[d(i-j)]^7}\right)\hat{n}_i \hat{n}_j .
\label{eq:expansion}
\end{aligned}
\end{equation}
Here, the second term represents a coupling between the spins and the atomic vibrations, commonly referred to as spin-phonon coupling. This result indicates that the dominant correction arises from motion along the $x$-direction, while contributions from motion in the transverse directions are negligible. To derive the effective spin-spin interactions induced by this spin-phonon coupling, we generalize the calculation of Ref.~\cite{PhysRevLett.131.093002} to the many-body case by performing a unitary transformation
\begin{equation}
    \hat{U}=\exp\Bigg({i\sum_{j=1}^{N}\sum_{i\neq j}\frac{6C_6 }{d^7(i-j)^7m\omega^2}\hat{n}_i \hat{n}_j\hat{p}_{x,j}}\Bigg).
    \label{eq:unitary}
\end{equation}
Physically, this corresponds to a spatial translation of each atom, with the displacement determined by the internal states of all atoms. From this expression, we also identify that the validity of the small-displacement approximation requires the condition $\text{max}\left(\frac{\delta x}{d}\right)\sim\frac{6C_6}{m \omega^2 d^8} \ll 1.$ The Hamiltonian now becomes:
\begin{equation}
  \begin{aligned}
    \hat{H} = &\sum_{j=1}^{N} \Bigg(\frac{\Omega}{2} \hat{U}\hat{\sigma}^x_j\hat{U}^\dagger -\Delta \hat{n}_j+\frac{\hat{\mathbf{p}}^2_j}{2m} + \frac{1}{2}m\omega^2\delta \hat{\mathbf{r}}^2_j\Bigg) \\&+\sum_{i<j}\frac{C_6\hat{n}_i \hat{n}_j}{d^6(j-i)^6} -\sum_{j=1}^N\frac{18C_6^2}{d^{14}m\omega^2}\hat{n}_j\Bigg(\sum_{k\neq 0}\frac{\hat{n}_{j+k}}{k^7}\Bigg)^2,
  \label{new hamiltonian_U}
  \end{aligned}
  \end{equation}
The last term represents multi-spin interactions generated by the spin-phonon coupling.

  \begin{figure*}[t]
    \centering
    \includegraphics[width=0.85\linewidth]{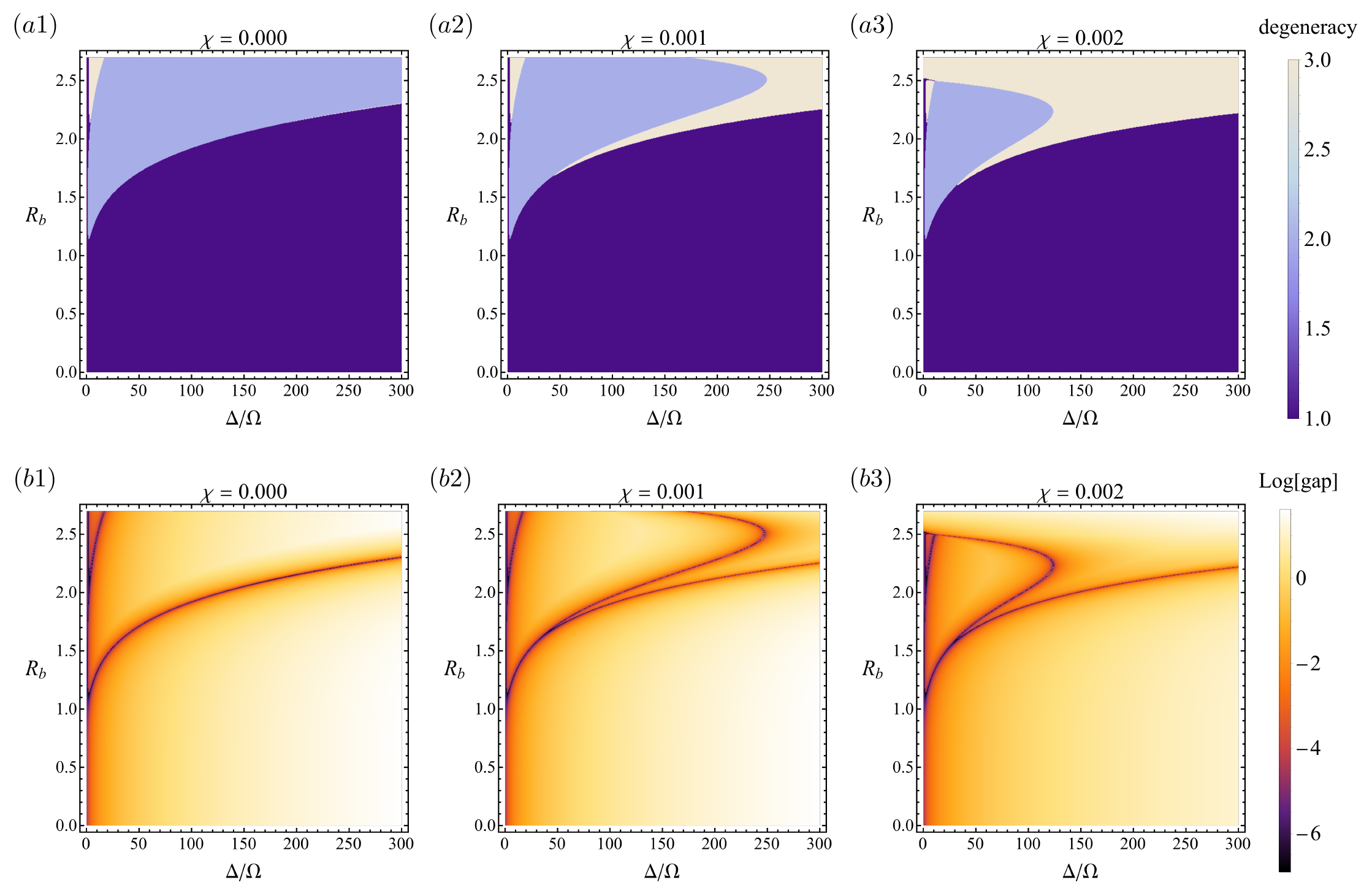}
    \caption{We present numerical results for the ground-state properties of the Hamiltonian~\eqref{eqn:Hmain} with $N = 12$ the periodic boundary condition, including panels (a1-a3) showing ground-state degeneracy and panels (b1-b3) showing the energy gap between the ground state and the first excited state. The results clearly demonstrate the emergence of a new phase with $\mathbb{Z}_3$ order, originating from the phonon-induced three-spin interactions.   }
    \label{fig:equilibrium}
  \end{figure*}

To further simplify the model, we introduce two additional approximations. First, we assume that the displacement is small not only compared to the interatomic spacing $d$, but also compared to the harmonic length $a_{\text{ho}}=1/\sqrt{m\omega}$: 
\begin{equation}\label{eqn:condition}
\text{max}\left(\frac{\delta x}{a_{\text{ho}}}\right)\sim\frac{6C_6}{m \omega^2 d^7 a_{\text{ho}}} \ll 1.
\end{equation}
This condition allows us to neglect the difference between $\hat{U}\hat{\sigma}^x_j\hat{U}^\dagger$ and $\sigma^x_j$. Physically, this is because the coupling between the ground and Rydberg states is generally reduced by the wavefunction overlap between their motional ground states. However, when condition \eqref{eqn:condition} is satisfied, the overlap remains close to unity. Under this approximation, the translated motional degrees of freedom become effectively decoupled from the spins and remain in their ground state. For conciseness, we will omit them from the Hamiltonian. Second, the last term in Eq.~\eqref{new hamiltonian_U} includes contributions from all values of $k$. However, the contributions from large $k$ decay rapidly. Therefore, we truncate the summation to $k=\pm 1$. We have verified that increasing the cutoff in $k$ does not affect the results presented in later sections. These approximations lead to an effective spin model
\begin{equation}
\begin{aligned}
    \hat{H} = &\sum_{j=1}^{N}  \bigg(\frac{\Omega}{2} \hat{\sigma}^x_j -\Delta \hat{n}_j\bigg) +\sum_{i<j}\frac{\Omega R_b^6 }{(j-i)^6}\hat{n}_i \hat{n}_j \\&-\sum_{j=1}^N{\chi \Omega R_b^{12}}\hat{n}_j(\hat{n}_{j-1}-\hat{n}_{j+1})^2.
\label{eqn:Hmain}
\end{aligned}
\end{equation}
This is the model that we focus on throughout this manuscript. Here, we introduce the dimensionless blockade radius $R_b$, defined in units of the lattice constant, via the condition $V(R_b)=\frac{C_6}{d^6R_b^6} \equiv \Omega$. In addition, we define the strength of the three-spin interaction as $\chi \equiv \frac{18\Omega}{m\omega^2 d^2}$. We focus on the weak Rabi coupling limit with fixed $R_b$, which corresponds to $\chi \ll 1$. In the following sections, we first analyze the model from both equilibrium and non-equilibrium perspectives, and then discuss its experimental realization.

\emph{ \color{blue}Equilibrium Phases.--} We begin by studying the ground-state phase diagram of the Hamiltonian~\eqref{eqn:Hmain}. Since $\chi \ll 1$, the phonon-induced interactions become significant only when $R_b^{12} \chi \sim \mathcal{O}(1)$. However, this condition implies a large van der Waals interaction between Rydberg atoms, which greatly exceeds the Rabi frequency $\Omega$. Therefore, we are focusing on the weak Rabi coupling regime. Additionally, we consider a large detuning $\Delta/\Omega$, which can compensate for the large energy scale introduced by the van der Waals interaction and enable non-trivial phase transitions. 

Our numerical results for $N = 12$ (with periodic boundary conditions) are presented in Fig.~\ref{fig:equilibrium}, where we show both the ground-state degeneracy and the energy gap between the ground state and the first excited state for different values of $\chi$. For $\chi = 0$, the model primarily exhibits three distinct phases~\cite{Bernien:2017ubn,2020NatPh..16..132B,Ebadi:2020ldi,Scholl:2020hzx}, each characterized by a different pattern of $\mathbb{Z}_{\text{N}}$ symmetry breaking:
(i) For sufficiently large $\Delta$, the system is in a symmetric phase with the fixed-point wavefunction $\ket{\uparrow\uparrow\uparrow\uparrow\cdots}$;
(ii) As $\Delta$ decreases, the system undergoes a transition into a $\mathbb{Z}_2$-ordered phase, with fixed-point wavefunction $\ket{\uparrow\downarrow\uparrow\downarrow\cdots}$;
(iii) For even smaller $\Delta$, the system enters the conventional $\mathbb{Z}_3$-ordered phase, characterized by the fixed-point wavefunction $\ket{\uparrow\downarrow\downarrow\uparrow\cdots}$. If the detunning $\Delta$ further decreases, $\mathbb{Z}_{\text{N}}$ orders with $N\geq 4$ can emerge. 

We now investigate the effect of phonons by introducing a small $\chi$. From both the ground-state degeneracy and the energy gap, we observe the emergence of a new phase characterized by three-fold ground-state degeneracy. This phase appears near the transition point between the fully polarized phase and the $\mathbb{Z}_2$ phase, a parameter regime where the van der Waals interaction is largely compensated by the detuning—consistent with our expectations. Moreover, as $\chi$ increases, the extent of this phase grows, confirming that it originates from the phonon-induced three-spin interaction. By directly analyzing the wavefunction, we identify the new phase as being described by the fixed-point wavefunction $\ket{\uparrow\uparrow\downarrow\uparrow\cdots}$, which corresponds to a $\mathbb{Z}_3$-ordered state with a spin configuration distinct from that of the conventional $\mathbb{Z}_3$ phase. 

  \begin{figure*}[t]
    \centering
    \includegraphics[width=0.98\linewidth]{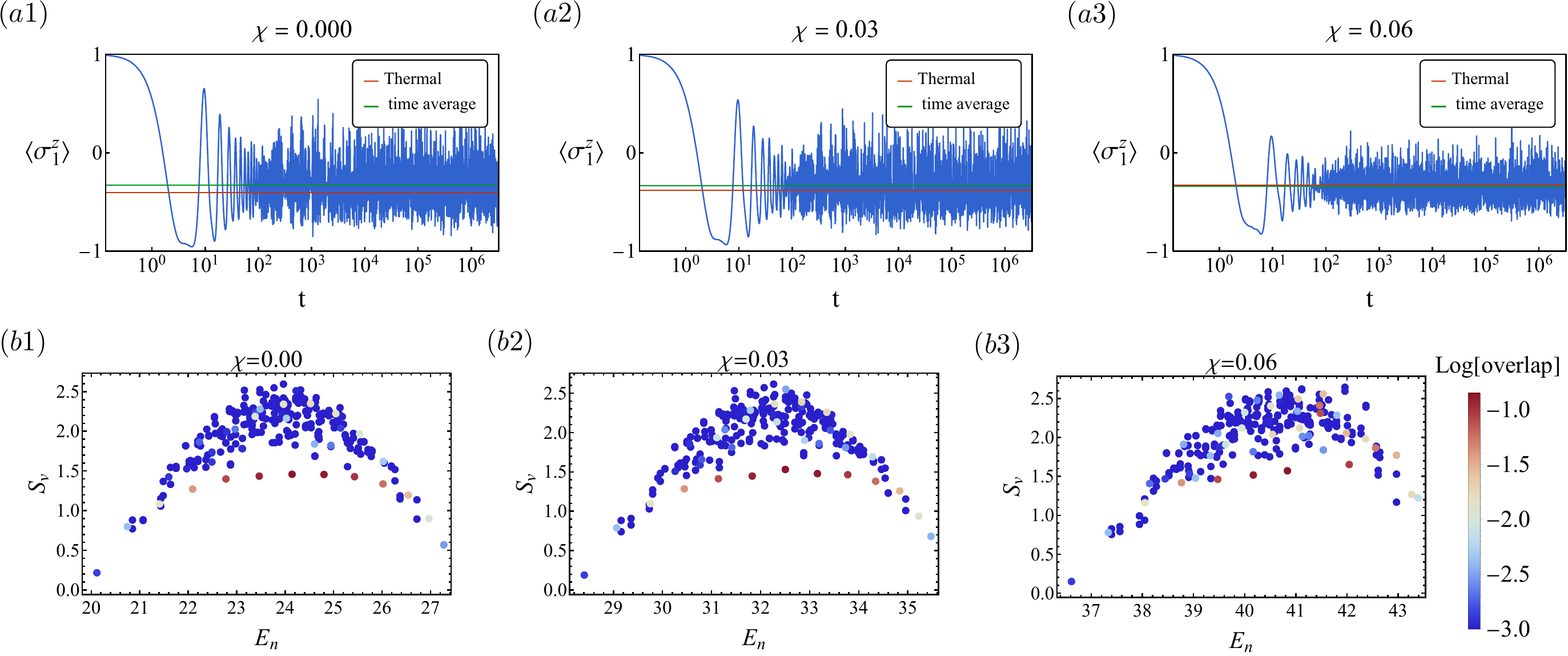}
    \caption{We present numerical results on the dynamical properties of the Hamiltonian~\eqref{eqn:Hmain} with $N = 12$ under periodic boundary conditions. Here, we set $\Omega=1$ as the energy unit. Panels (a1-a3) show the quench dynamics of the observable $\langle \sigma^z_1 \rangle$, while panels (b1-b3) display the many-body spectrum restricted to states with small nearest-neighbor Rydberg pair occupations, $\langle \sum_i \hat{n}_i \hat{n}_{i+1} \rangle < 0.5$, along with the overlap between each eigenstate and the $\mathbb{Z}_2$ state. These results suggest that spin-phonon interactions suppress the violation of quantum thermalization.}
    \label{fig:nonequilibrium}
  \end{figure*}

We provide additional arguments for the emergence of this phase. First, by examining each nearest-neighbor spin pair, we observe that only the states $\ket{\uparrow\uparrow}$, $\ket{\uparrow\downarrow}$, and $\ket{\downarrow\uparrow}$ appear in the fixed-point wavefunction. These configurations are characteristic of the fully polarized phase or the $\mathbb{Z}_2$ phase, and therefore span the effective low-energy Hilbert space near their transition point. This situation is closely analogous to the traditional derivation of the PXP model~\cite{Turner_2018,Turner:2018yco}, which focuses on the regime where $V(1) \gg \Omega$ and $V(2) \ll \Omega$, identifying the states $\ket{\downarrow\downarrow}$, $\ket{\uparrow\downarrow}$, and $\ket{\downarrow\uparrow}$ as the low-energy subspace. Consequently, our regime is related to the traditional PXP regime by a spin reflection $\ket{\uparrow} \leftrightarrow \ket{\downarrow}$. When the next-to-nearest-neighbor interaction is negligible, it can also be captured by the effective PXP-like Hamiltonian
\begin{equation}
\hat{H}_{\text{eff}}= \Omega\sum_{j=1}^N\left(\hat{P}\sigma^x_j\hat{P}-\chi R_b^{12}\hat{n}_j(\hat{n}_{j-1}-\hat{n}_{j+1})^2\right).
\end{equation}
Here, the projector $\hat{P}$ excludes any state containing two neighboring atoms both in the ground state. In the PXP model, adding next-to-nearest-neighbor repulsive interactions between $\uparrow$ spins leads to the conventional $\mathbb{Z}_3$ order with fixed-point wavefunction $\ket{\uparrow\downarrow\downarrow\uparrow\cdots}$. In our model, the new $\mathbb{Z}_3$ phase arises similarly, up to an additional spin reflection. Direct analysis also confirms that our configuration $\ket{\uparrow\uparrow\downarrow\uparrow\cdots}$ is favored by the three-spin interaction, which requires that each Rydberg atom be surrounded by one ground-state atom and another Rydberg atom. 

\emph{ \color{blue}Non-equilibrium Dynamics.--} Next, we investigate how the spin-phonon interaction modifies the dynamical properties of the system. We set $\Delta=0$, ${R_b}=1.3$, and $N=12$ in the Hamiltonian \eqref{eqn:Hmain}, where the system with $\chi = 0$ has a low-energy subspace described by the standard PXP model, which exhibits quantum many-body scar states. These scar states are non-thermal states embedded within an otherwise thermalizing spectrum. We focus on two distinct characteristics of these non-thermal states~\cite{Turner_2018,PhysRevB.105.125123}. 
\begin{enumerate}
\item We study the quench dynamics of the system starting from a $\mathbb{Z}_2$ initial state, $\ket{\Psi(0)} = \ket{\uparrow\downarrow\uparrow\downarrow\cdots}$. Evolving this state under the Hamiltonian~\eqref{eqn:Hmain} to time $t$, we compute the long-time average of the observable $\langle\hat{\sigma}^{z}_i\rangle$ as
\begin{equation}
    \bar{\sigma}^{z}_i = \lim_{T\rightarrow \infty}\lim_{t_0\rightarrow \infty}\frac{1}{T} \int_{t_{0}}^{t_{0}+T}\langle\Psi(t)| \hat{\sigma}_{i}^{z}|\Psi(t)\rangle d t ,
\end{equation}
where $|\Psi(t)\rangle=e^{-i\hat{H}t}|\Psi(0)\rangle$. We compare this with the thermal equilibrium expectation value~\cite{PhysRevA.43.2046,PhysRevE.50.888,Rigol:2007juv,DAlessio:2015qtq}
\begin{equation}
\sigma^{z}_{\mathrm{th}} = \mathrm{Tr}(\rho_{\mathrm{th}} \hat{\sigma}^z_i),
\end{equation}
where the thermal density matrix is given by $\rho_{\mathrm{th}} = e^{-\beta \hat{H}} / Z$, with the partition function $Z = \mathrm{Tr}(e^{-\beta \hat{H}})$.
The effective inverse temperature $\beta$ is determined by matching the total energy $\mathrm{Tr}(\rho_{\mathrm{th}} \hat{H})=\langle \Psi(0)|\hat{H}|\Psi(0)\rangle$.

\item We also compute the bipartite entanglement entropy of the eigenstates $|E_n\rangle$ for a subsystem $v$ consisting of $N_v = 6$ contiguous sites. Specifically, we partition the full system into $v \cup \bar{v}$ and calculate the von Neumann entropy $S_v=-\mathrm{Tr}_{v}\left(\rho_v\ln \rho_v\right)$ where the reduced density matrix is given by $\rho_v=\mathrm{Tr}_{\bar{v}} |E_n\rangle \langle E_n|$. We plot the relation between the eigenenergy $E_n$ and the corresponding entanglement entropy $S_v$ to identify low-entanglement states at finite energy density, which are associated with scar states. For conciseness, we restrict our analysis to eigenstates with small nearest-neighbor Rydberg pair occupations, $\langle \sum_i \hat{n}_i \hat{n}_{i+1} \rangle < 0.5$, effectively selecting the low-energy subspace of the PXP model.
Additionally, to connect with the quench dynamics, we present the overlap between each eigenstate and the $\mathbb{Z}_2$ state $\ket{\Psi(0)} = \ket{\uparrow\downarrow\uparrow\downarrow\cdots}$.

\end{enumerate}

The numerical results are presented in Fig.~\ref{fig:nonequilibrium}. For $\chi = 0$, the quench dynamics starting from the $\mathbb{Z}_2$ initial state exhibits persistent oscillations of $\langle\hat{\sigma}^{z}_1\rangle$ over long times. The long-time average $\bar{\sigma}^{z}_1$ deviates from the thermal ensemble prediction, indicating a violation of the eigenstate thermalization hypothesis. Simultaneously, the entanglement entropy plot reveals a series of scar states that lie outside the main branch of thermalized states. These scar states provide significant support for the $\mathbb{Z}_2$ state. This behavior is consistent with previous numerical studies based on the PXP model~\cite{Turner_2018,PhysRevB.105.125123}. Upon introducing a small spin-phonon interaction, we find that the amplitude of the oscillations gradually decreases as $\chi$ increases. Moreover, the deviation between the thermal expectation value $\sigma^{z}_{\mathrm{th}}$ and the long-time average $\bar{\sigma}^{z}_1$ also diminishes for small values of $\chi$. These results indicate a gradual restoration of quantum thermalization as $\chi$ increases. To further support this observation, we analyze the entanglement entropy of the eigenstates. With a small but finite $\chi$, the scar states progressively approach the thermalized branch and nearly merge with it at $\chi \sim 0.06$. Additionally, the $\mathbb{Z}_2$ state begins to have greater overlap with eigenstates that possess higher entanglement. Both features in the spectrum are consistent with the trends observed in the quench dynamics~\cite{PhysRevLett.131.020402,PhysRevB.105.125123,Turner_2018}.

\emph{ \color{blue}Experimental consideration.--} Finally, we estimate the experimental parameters required to observe the spin-phonon effect. First, we consider the realization of the novel $\mathbb{Z}_3$ phase in thermal equilibrium. As an example, we take $\chi = 0.001$, consistent with FIG.\ref{fig:equilibrium}(a2). For a lattice spacing of $d = 3 \mu\text{m}$ and $35S$ Rydberg states of $^{39}\text{K}$ atoms, the van der Waals interaction strength is $C_6/d^6 \approx 0.214$ MHz~\cite{2017CoPhC.220..319S}. Assuming a weak Rabi frequency $\Omega = 9.5$ kHz and an oscillation frequency $\omega = 2\pi \times 68$ kHz~\cite{PhysRevLett.131.093002}, we obtain $\chi = 0.001$ and $R_b = 1.7$, placing the system within the parameter regime that supports the novel $\mathbb{Z}_3$ phase. Moreover, the condition in Eq.\eqref{eqn:condition} is satisfied, as ${6C_6}/{m \omega^2 d^7 a_{\text{ho}}} \sim 0.15$, validating the parameter regime considered in our study. It is also straightforward to verify that the parameters used in the non-equilibrium dynamics can be realized in a similar manner.

\emph{ \color{blue}Discussions.--} 
We investigate the impact of atomic vibrations on the quantum many-body physics of one-dimensional Rydberg atom arrays by incorporating spin-phonon coupling arising from the motion of atoms in optical tweezers. Going beyond the conventional frozen-atom approximation, we show that phonon-mediated interactions generate novel effective three-spin couplings, leading to the emergence of a new $\mathbb{Z}_3$ symmetry-breaking phase at thermal equilibrium. Additionally, we demonstrate that spin-phonon coupling suppresses the characteristic nonthermal behavior of $\mathbb{Z}_2$ quantum many-body scar states, thereby promoting thermalization. Our results highlight the significant role of motional degrees of freedom in shaping both equilibrium and dynamical quantum phases and could be readily explored in current experimental setups with Rydberg atom arrays.

We offer a few comments on potential directions for future research. While the present study focuses primarily on the perspective of quantum many-body physics, the Rydberg atom array also serves as an ideal platform for quantum simulation and quantum computation~\cite{RevModPhys.82.2313,2010NatPh...6..382W,Labuhn_2016,Bernien:2017ubn,2020NatPh..16..132B,Ebadi:2020ldi,Scholl:2020hzx,Wu:2020axb,Semeghini:2021wls,fang2024probingcriticalphenomenaopen,Cheng:2024nep,Bluvstein_2021,Morgado:2020jfo,Bluvstein:2021jsq,Ma:2023ltx,Scholl:2023cjt,Singh:2022qfv,Bluvstein:2023zmt,zhang2025observationnearcriticalkibblezurekscaling}. First, it would be interesting to explore whether the spin-phonon interactions in Rydberg arrays can be engineered to directly simulate the spin-phonon couplings found in solid-state systems. Such a development could provide valuable insights and enable advanced quantum simulations of strongly correlated materials. Second, the potential role of phonon degrees of freedom in quantum information processing merits thorough investigation. Specifically, one could examine whether phonons can serve as robust quantum memory elements or even as mediators of quantum logic gates. These avenues could expand the versatility of Rydberg atom platforms in quantum technologies.

\vspace{5pt}
\emph{ \color{blue}Acknowledgement.--} 
We thank Chengshu Li, Zeyu Liu, Tom Manovitz, Ning Sun, Yuke Zhang, and Shuyan Zhou for helpful discussions. This project is supported by the Shanghai Rising-Star Program under grant number 24QA2700300, the NSFC under grant 12374477, and the Innovation Program for Quantum Science and Technology 2024ZD0300101.

\bibliography{main.bbl}

\end{document}